\documentclass[12pt]{iopart}
\usepackage{iopams}
\usepackage{graphicx}
\usepackage[T1]{fontenc}
\def\sn{\mathop{\rm sn}\nolimits}
\def\cn{\mathop{\rm cn}\nolimits}
\def\dn{\mathop{\rm dn}\nolimits}
\begin{document}

\title{Generalised asymptotic solutions for the inflaton in the oscillatory phase of reheating}
\author{Gabriel \'Alvarez$^1$, Luis Mart\'{\i}nez Alonso$^1$ and Elena Medina$^2$}
\address{$^1$ Departamento de F\'{\i}sica Te\'orica,
                        Facultad de Ciencias F\'{\i}sicas,
                        Universidad Complutense,
                        28040 Madrid, Spain}
\address{$^2$ Departamento de Matem\'aticas,
                        Facultad de Ciencias,
                        Universidad de C\'adiz,
                        11510 Puerto Real, C\'adiz, Spain}

\begin{abstract}
We determine generalised asymptotic solutions for the inflaton field, the Hubble parameter, and the
equation-of-state parameter valid during the oscillatory phase of reheating for potentials that close
to their global minima behave as even monomial potentials. For the quadratic potential we derive
a generalised asymptotic expansion for the inflaton
with respect to the scale set by inverse powers of the cosmic time. For the quartic potential we
derive an explicit, two-term generalised asymptotic solution in terms of Jacobi elliptic functions,
with a scale set by inverse powers of the square root of the cosmic time.
Finally, in the general case, we find similar two-term solutions where the leading order term is
defined implicitly in terms of the Gauss' hypergeometric function. The relation between the leading terms
of the instantaneous equation-of-state parameter and different averaged values is discussed in the general case.
\end{abstract}

%
%
%
\section{Introduction\label{sec:intro}}
The reheating period in inflationary cosmology is a transition period connecting the end of inflation to the
radiation-dominated era~\cite{AS82,KA94,ST95,KA97,MU05,BA06,BA09}. During reheating,
the energy of the inflaton field $\Phi$
transforms into relativistic particles. A basic model for the inflaton decay is given by the system of
equations~\cite{AS82,TU83,MA10}
\begin{equation}
	\label{eq:id1}
	\ddot{\Phi}+(3H+\Gamma)\dot{\Phi}+V'(\Phi)=0,
\end{equation}
\begin{equation}
	\label{eq:id2}
	\dot{\rho}_{\gamma}=-4 \rho_{\gamma}H+(1+w_{\Phi})\Gamma \rho_{\Phi},
\end{equation}
where dots denote derivatives with respect to the cosmic time $t$ and
$V(\Phi)$ is the inflaton potential. The inflaton pressure, energy density and
equation of state (eos) parameter are given respectively by
\begin{equation}
	\label{eq:pf}
	p_{\Phi}=\frac{1}{2} \dot{\Phi}^2 -V(\Phi),
\end{equation}
\begin{equation}
	\label{eq:rf}
	\rho_{\Phi}=\frac{1}{2} \dot{\Phi}^2 +V(\Phi),
\end{equation}
\begin{equation}
	w_{\Phi}=p_{\Phi}/ \rho_{\Phi},
\end{equation}
$\Gamma$ is the total inflaton decay rate, $\rho_{\gamma}$ is  energy density
of the radiation, and the  Hubble parameter $H$ satisfies the Friedmann-Lema\^{\i}tre equation
\begin{equation}
	\label{eq:h}
	H^2=\frac{1}{3\mathrm{M_{Pl}}^2}(\rho_{\Phi}+\rho_{\gamma}),
\end{equation}
with $\mathrm{M_{Pl}}=2.435\times 10^{18}\mathrm{GeV}$ being the reduced Planck mass.

The model~(\ref{eq:id1})--(\ref{eq:id2}) represents reheating in terms of an effective fluid with total 
pressure $p=p_{\Phi}+p_{\gamma}$ and density $\rho=\rho_{\Phi}+\rho_{\gamma}$.
The radiation eos $p_{\gamma}=\rho_{\gamma}/3$ leads to an eos parameter during reheating given by~\cite{MA10}
 \begin{equation}\label{wre}
 	w_\mathrm{re}=\frac{p}{\rho}=\frac{ \frac{1}{2} \dot{\Phi}^2 -V(\Phi)+\frac{1}{3}\rho_{\gamma}}{\frac{1}{2} \dot{\Phi}^2 +V(\Phi)+\rho_{\gamma}}.
 \end{equation}

The first phase of reheating can be described as an oscillatory phase around the true minimum of the inflaton
potential, wherein the inflaton is not yet effectively coupled to the radiation fields and  the friction term 
$\Gamma\dot{\Phi}$ in equation~(\ref{eq:id1}) can be neglected. During this period, equations~(\ref{eq:id1}) and~(\ref{eq:h}) reduce to
\begin{equation}
 	\label{eq:ids1}
	\ddot{\Phi}+3H\dot{\Phi}+V'(\Phi)=0,
\end{equation}
\begin{equation}
	\label{eq:hs}
	H^2=\frac{1}{3\mathrm{M_{Pl}}^2}\left(\frac{1}{2} \dot{\Phi}^2 +V(\Phi)\right).
\end{equation}

Conversely, in the last phase of reheating the term $3H\dot{\Phi}$ can be neglected with respect to $\Gamma\dot{\Phi}$
in~(\ref{eq:id1}), leading to the simple equation
\begin{equation}
	\label{eq:rp2}
	\ddot{\Phi}+\Gamma\dot{\Phi}+V'(\Phi)=0.
\end{equation}

During the oscillatory period, where $\rho_{\gamma}\ll \rho_{\Phi}$, $w_\mathrm{re}$ reduces to $w_\Phi$, whereas at the end of reheating
$\rho\sim \rho_{\gamma}$ and $w_\mathrm{re}$ tends to $1/3$, i.e., to the onset of the radiation era.

The aim of this paper is to determine generalised asymptotic solutions for the inflaton field $\Phi(t)$ and derived quantities
valid during the oscillatory phase of reheating for potentials that close to their global minima behave as the even monomial potentials 
\begin{equation}
	\label{eq:v}
	V(\Phi)=\frac{M^4}{p}\left(\frac{\Phi}{\mathrm{M_{Pl}}}\right)^{2p},\quad (p=1,2,\ldots).
\end{equation}

Turner~\cite{TU83}, Mukhanov~\cite{MU05}, and Rendall~\cite{RE07} emphasise that during the oscillatory phase
the model is approximated \emph{in some averaged sense} by a perfect fluid with a linear equation of state.

Since these generalised asymptotic solutions will be formally valid as the cosmic time $t$ tends to
infinity, it may be raised the question of its physical relevance in comparison with the exact solutions
of the full equations~(\ref{eq:id1}),~(\ref{eq:id2}) and~(\ref{eq:h}),
because the power-law-damped oscillatory approximation will certainly break down when
$H\sim\Gamma$ and $\rho_{\gamma}\gtrsim\rho_{\Phi}$, and the oscillations become exponentially damped.
However, typical values of the parameters for, e.g., the nondegenerate case $p=1$,
are $M=3\times10^{-3}\mathrm{M_{Pl}}$ and $\Gamma=1.375\times 10^9\mathrm{ GeV}$, and order-of-magnitude
estimates and numerical calculations confirm that there is a large time interval (several e-folds) in which the asymptotic 
solutions in the oscillatory phase will be valid (see, for example, FIG.~13 in reference~\cite{MA10}).

If we write
\begin{equation}
	 	\Phi(t) = \sqrt{\frac{2}{3}}\,\mathrm{M_{Pl}}\,\varphi(\tau),
\end{equation}
where
\begin{equation}
	\label{eq:m}
	m = \sqrt{\frac{3}{p}} \left(\frac{2}{3}\right)^{p/2} \frac{M^2}{\mathrm{M_{Pl}}},
\end{equation}
and
\begin{equation}
	\label{eq:tau}
	\tau = m t,
\end{equation}
and substitute equations~(\ref{eq:hs}) and~(\ref{eq:v}) into~(\ref{eq:ids1}), we find that the system is equivalent
to the single non-linear differential equation
\begin{equation}
	\label{eq:me}
	\varphi''(\tau) + ( \varphi'(\tau)^2 + \varphi(\tau)^{2p} )^{1/2} \varphi'(\tau) + p \varphi(\tau)^{2p-1} = 0.
\end{equation}
Likewise, the Hubble parameter can be written in terms of $\varphi$ as
\begin{equation}
	\label{eq:ht}
	H(t) = \frac{m}{3} h(\tau),
\end{equation}
where for later convenience we have introduced the reduced Hubble parameter
\begin{equation}
	\label{eq:htau}
	h(\tau) = (\varphi'(\tau)^2+ \varphi(\tau)^{2p})^{1/2}.
\end{equation}
Incidentally, by differentiation the former equation and using~(\ref{eq:me}) we find that
\begin{equation}
	\label{eq:hprimetau}
	h'(\tau) = -\varphi'(\tau)^2.
\end{equation}
As usual, primes denote derivatives of $\varphi$ with respect to its argument.

By introducing polar coordinates in the $(\varphi,\varphi')$ plane, Mukhanov~\cite{MU05} and Rendall~\cite{RE07}
find the first two terms  of the asymptotic expansion of the inflaton field for the quadratic potential corresponding to
$p=1$ in equation~(\ref{eq:v}). (In fact, equation~(24) of reference~\cite{RE07} corrects a mistake in equation~(5.45) of
reference~\cite{MU05}.) For general values of $p$ and using different averaging procedures, the authors
arrive at a common constant leading term for the averaged value
\begin{equation}
	\label{eq:winf}
	\langle w_\Phi \rangle \sim \frac{p-1}{p+1},\quad (t\to\infty),
\end{equation}
i.e.,~to the above-mentioned averaged linear equation of state. Thus, $\langle w_\Phi \rangle\sim 0$ for $p=1$
and $\langle w_\Phi \rangle\sim 1/3$ for $p=2$, which correspond to a universe dominated by
nonrelativistic matter and to a universe dominated by radiation respectively. It was noted by Turner~\cite{TU83}
that the amplitude of the inflaton oscillations decreases as $\tau^{1/p}$. However, for $p\geq 2$ we are not aware
of any explicit results for the asymptotic form of the inflaton field.

In section~\ref{sec:v2} we reconsider the nondegenerate harmonic potential discussed by Mukhanov~\cite{MU05}
and Rendall~\cite{RE07}. First we present a method to calculate a generalised asymptotic expansion for the
inflaton field
\begin{equation}
	\label{eq:ans}
	\varphi(\tau) \sim \sum_{n=1}^\infty \frac{\varphi_n(\tau)}{\tau^{n}},\quad (\tau\to\infty),
\end{equation}
where the coefficients $\varphi_n$ are periodic (and therefore bounded) functions. To calculate these expansions we substitute
the ansatz~(\ref{eq:ans}) into equation~(\ref{eq:me}) and solve recursively the system of differential
equations for the $\varphi_n$. The requisite that $\varphi_n$ be bounded is implemented by imposing
the cancellation of resonant terms in these equations. As a byproduct, we will show how to calculate
asymptotic expansions for derived quantities like $H$ and $w_\Phi$.
A formal proof of the existence of the generalised asymptotic expansion~(\ref{eq:ans}) is deferred to appendix~A.

A straightforward generalisation of the ansatz~(\ref{eq:ans}) does not work for potentials~(\ref{eq:v}) with $p>1$
because the resonant terms cannot be systematically eliminated past a certain order.
Therefore we look for generalised asymptotic formulas with a finite number of
terms~\cite{MI06}. In section~\ref{sec:v4} we discuss separately the quartic potential.
This is a particularly relevant potential because the corresponding oscillatory phase of reheating
is radiation-dominated and the universe would be radiation-dominated since the end of inflation,
with an ensuing reduction of the uncertainty on the period were the observable perturbations
of the universe were produced~\cite{LI03}. We obtain explicit asymptotic formulas for the inflaton field
and derived quantities in terms of Jacobi elliptic functions. In section~\ref{sec:vp}
we give a general derivation valid for any value of $p$, where the leading term of the asymptotic formula for the inflaton
field is defined implicitly in terms of Gauss' hypergeometric function, whereas the second term can be given explicitly
in terms of the first. We also obtain the leading term of the instantaneous eos parameter $w_{\Phi}$ and show how
different averaged values of  $w_{\Phi}$ coincide with the standard result (\ref{eq:winf}).
The paper ends with some brief remarks.
\section{The harmonic potential\label{sec:v2}}
In this section we consider the nondegenerate harmonic potential discussed by Turner~\cite{TU83}, Mukhanov~\cite{MU05},
and Rendall~\cite{RE07}, which corresponds to setting $p=1$ in equation~(\ref{eq:v}),
and for which equation~(\ref{eq:me}) reduces to
\begin{equation}
	\label{eq:mep1}
	\varphi''(\tau) + ( \varphi'(\tau)^2 + \varphi(\tau)^{2} )^{1/2} \varphi'(\tau) + \varphi(\tau) = 0.
\end{equation}

As we mentioned in the introduction, we look for a generalised asymptotic expansion of $\varphi(\tau)$ with
respect to the asymptotic sequence $\tau^{-n}$, $(n=1,2,3,\ldots)$ (see section 10.3 of~\cite{OL97}),
\begin{equation}
	\label{eq:ansp1}
	\varphi(\tau)\sim\sum_{n=1}^\infty \frac{\varphi_n(\tau)}{\tau^n},\quad (\tau\to\infty)
\end{equation}
where the $\varphi_n(\tau)$ are periodic functions.
Substituting equation~(\ref{eq:ansp1}) into~(\ref{eq:mep1}) and arranging the result by inverse powers of $\tau$
(note that all the derivatives of the $\varphi_n(\tau)$ will also be periodic and therefore bounded), we find
an infinite system of differential equations for the $\varphi_n(\tau)$. The first three, which are enough to show the
pattern, are:
\begin{equation}
	\label{eq:p1f1}
	\varphi''_1 + \varphi_1 = 0,
\end{equation}
\begin{equation}
	\label{eq:p1f2}
	\varphi''_2 + \varphi_2 = (2 - ( (\varphi'_1)^2 + \varphi_1^{2} )^{1/2})\varphi'_1,
\end{equation}
\begin{equation}
	\label{eq:p1f3}
	\varphi''_3 + \varphi_3 = 4 \varphi'_2 - 2 \varphi_1 -
	\frac{\varphi _1^2 \varphi' _2-2 \varphi _1 (\varphi' _1)^2+\varphi _2 \varphi _1 \varphi' _1
	        +2 (\varphi' _1)^2 \varphi'_2-\varphi _1^3}{( (\varphi'_1)^2 + \varphi_1^{2} )^{1/2}}.
\end{equation}

Equation~(\ref{eq:p1f1}) is a harmonic oscillator, whose general solution can be written as
\begin{equation}
	\varphi_1(\tau) = b \cos(\tau-\tau_0).
\end{equation}
Substituting this solution into the right hand side of equation~(\ref{eq:p1f2}) we find
\begin{equation}
	\varphi''_2 + \varphi_2 = b(b-2)\sin(\tau-\tau_0).
\end{equation}
Note that the inhomogeneous term is resonant. Therefore, the solution for $\varphi_2(\tau)$ would be
unbounded unless we set $b=2$ ($b=0$ leads to a trivial solution for $\varphi_1(\tau)$),
and we have again a harmonic oscillator with general solution
\begin{equation}
	\varphi_2(\tau) = c \cos(\tau-\tau_0) + d \sin(\tau-\tau_0).
\end{equation}
Substituting this solution into the right-hand side of equation~(\ref{eq:p1f3}) we find,
\begin{equation}
	\varphi''_3 + \varphi_3 =  \left(1+2d\right) \cos (\tau-\tau_0)-\cos(3(\tau-\tau_0)).
\end{equation}
The first term in the right-hand side is again resonant, and the solution for $\varphi_3(\tau)$
would be unbounded unless we set $d=-1/2$. The coefficient $c$ remains free.

This recursive procedure can be easily programmed in a computer: at each stage the
function $\varphi_n(\tau)$ is substituted into the right-hand side of the equation for $\varphi_{n+1}(\tau)$
and resonant terms eliminated. It turns out that all the coefficients depend only on $\tau_0$ and $c$.
The first four coefficients are:
\begin{eqnarray}
	\varphi_1(\tau) = & 2 \cos (\tau-\tau_0),\label{eq:2}\\
	\varphi_2(\tau) = & c \cos (\tau-\tau_0) - \frac{1}{2} \sin (\tau-\tau_0),\\
	\varphi_3(\tau) = & \frac{1}{16} \left(8 c^2+9\right) \cos (\tau-\tau_0)-\frac{c}{2}  \sin (\tau-\tau_0)+\frac{1}{8} \cos (3 (\tau-\tau_0)),\\
	\varphi_4(\tau) = & \frac{c}{32} \left(8 c^2+27\right) \cos (\tau-\tau_0)-\frac{1}{192} \left(72 c^2+7\right) \sin (\tau-\tau_0)\\
	                      &{}+\frac{3c}{16} \cos (3 (\tau-\tau_0))-\frac{5}{32} \sin (3 (\tau-\tau_0)).
\end{eqnarray}
In light of these four coefficients one might conjecture the general form
\begin{equation}
	\label{eq:fn}
 	\varphi_n(\tau) = \sum_{k=1,k\,\mathrm{odd}}^{n} c_{n,k} \, \cos (k (\tau-\tau_0)) + d_{n,k} \sin(k(\tau-\tau_0)),
\end{equation}
with $d_{n,n}=0$ for $n$ odd. (In fact, the most efficient way to calculate the expansion is to use this ansatz
and to equate coefficients of the inverse powers of $\tau$.) A formal proof of the fact that equation~(\ref{eq:fn})
is indeed true is deferred to Appendix~A.

\begin{figure}
	 \begin{center}
		\includegraphics[width=10cm]{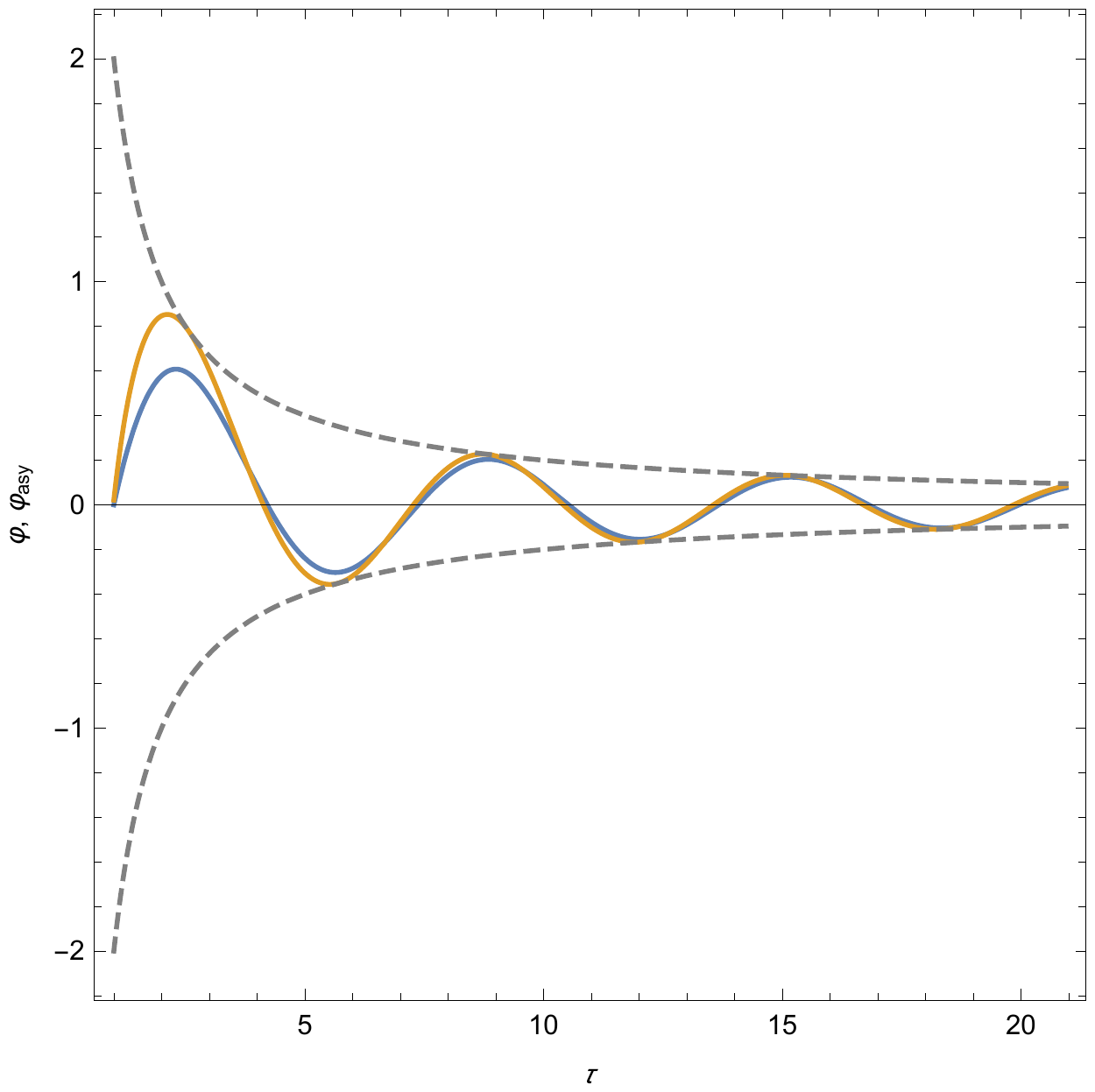}
	\end{center}
	\caption{Numerical integration (blue line) with initial conditions $\varphi(1)=0$, $\varphi'(1)=1$ and leading term
	              of the asymptotic expansion (brown line) for the differential equation~(\ref{eq:mep1}) corresponding
	              to the quadratic potential. The dashed lines are the envelopes $\pm 2/\tau$ that
	              set the scale for the generalised asymptotic expansion~(\ref{eq:ansp1}).\label{fig:v2s}}
\end{figure}

To illustrate the accuracy of this generalised asymptotic solution, in figure~(\ref{fig:v2s}) we plot the results of 
a numerical integration of the differential equation~(\ref{eq:mep1}) with initial conditions $\varphi(1)=0$, $\varphi'(1)=1$,
as well as the leading term of the asymptotic expansion~(\ref{eq:ansp1}). The numerical solution quickly
tends to the asymptotic solution. The dashed lines are the envelopes $\pm 2/\tau$ that
set the scale for the generalised asymptotic expansion~(\ref{eq:ansp1}).
(The factor $2$ is precisely the $2$ in equation~(\ref{eq:2}) for $\varphi_1(\tau)$.) Other initial conditions
lead to similar plots and are not represented to avoid clutter in the figure.

As a consequence of the main expansion~(\ref{eq:ansp1}) we can calculate the corresponding expansion for
the reduced Hubble parameter~(\ref{eq:htau}),
\begin{equation}
	\label{eq:htaup1}
	h(\tau) \sim \sum_{n=1}^\infty\frac{h_n(\tau)}{\tau^{n}},
\end{equation}
whose first terms are
\begin{eqnarray}
	h_1(\tau) =& 2,\label{eq:h1p1}\\  
	h_2(\tau) =& c+\sin(2(\tau-\tau_0)),\\  
	h_3(\tau) =&\frac{1}{4}(3+2c^2)+\frac{3}{2}\cos(2(\tau-\tau_0))+c\sin(2(\tau-\tau_0)).
\end{eqnarray}

\begin{figure}
	 \begin{center}
		\includegraphics[width=10cm]{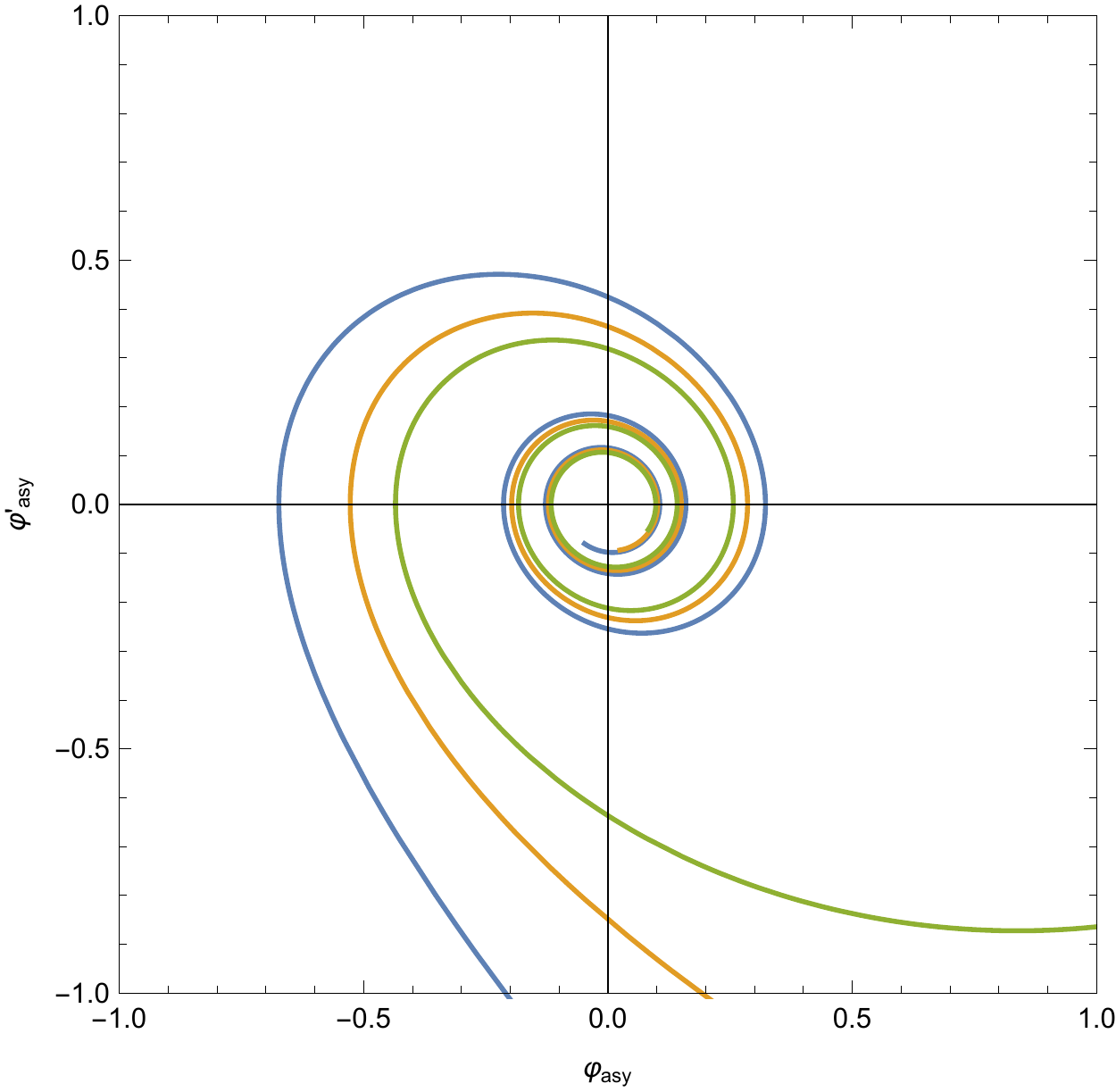}
	\end{center}
	\caption{Three typical asymptotic trajectories on the phase map of the differential equation~(\ref{eq:mep1})
	              corresponding to the quadratic potential. Note that for sufficiently large times the approximation
	              of the representative point to the origin is monotonic.\label{fig:v2p}}
\end{figure}

In figure~\ref{fig:v2p} we plot three typical asymptotic trajectories on the phase map of the differential
equation~(\ref{eq:mep1}). Note that for sufficiently large times the approximation of the representative point
to the origin is monotonic: in fact, for the quadratic potential the distance of the representative point to the origin
of the phase plane is just the reduced Hubble constant $h(\tau)\sim 2/\tau$, as shown in equation~(\ref{eq:h1p1}).

Using equation~(\ref{eq:ht}) we find the expansion for the Hubble parameter
\begin{eqnarray}
	H(t) \sim & \frac{2}{3t} + \frac{c+\sin(2m(t-t_0))}{3 m t^2} \nonumber\\
	               & {}+ \frac{3+2 c^2+4 c \sin (2 m(t-t_0))+6 \cos (2m (t-t_0))}{12 m^2t^3}\cdots.
\end{eqnarray}
Here $m$ is the value of equation~(\ref{eq:m}) for $p=1$, i.e, $m=\sqrt{2} M^2/M_\mathrm{Pl}$, and $\tau_0 = m t_0$.
Note that in this case the inflaton potential reduces to
\begin{equation}
	V(\Phi) = \frac{1}{2} m^2 \Phi^2,
\end{equation}
i.e., $m$ is the $p=1$ inflaton mass. Likewise, using
\begin{equation}
	\label{esp}
	w_\Phi(t) = -1 -\frac{2}{3} \frac{\dot{H}}{H^2},
\end{equation}
we find the expansion for the equation of state parameter
\begin{eqnarray}
\fl
	w_\Phi(t)
	\sim & -\cos(2m(t-t_0))+\frac{3\sin(2m(t-t_0))+\sin(4m(t-t_0))}{2mt}
	\nonumber\\
\fl
	& {}+ \frac{1}{16m^2 t^2} \Big( 4+23 \cos(2m(t-t_0))+12 c \sin (2m(t-t_0))
	\nonumber\\
\fl
	&\qquad{}+20 \cos (4m(t-t_0))+4c \sin(4m(t-t_0))+3 \cos(6m(t-t_0))\Big) + \cdots .
	\label{eq:wf2t}
\end{eqnarray}
Note that these first two terms of the complete asymptotic expansion for $w_\Phi(t)$ are independent of $c$.
Note also that the average of the leading term is trivially zero.
\section{The quartic potential\label{sec:v4}}
In this section we discuss the quartic potential, which corresponds to setting $p=2$ in equation~(\ref{eq:me}):
\begin{equation}
	\label{eq:mep2}
	\varphi''(\tau) + ( \varphi'(\tau)^2 + \varphi(\tau)^4 )^{1/2} \varphi'(\tau) + 2 \varphi(\tau)^3 = 0.
\end{equation}
As was noted by Turner~\cite{TU83} and numerical calculations confirm, the solutions of this differential equation oscillate
with a maximal amplitude that decreases as $1/\tau^{1/2}$. Therefore, one might try a generalised asymptotic solution similar
to equation~(\ref{eq:ansp1}) but with $\tau$ replaced by $\tau^{1/2}$ on the right-hand side of that equation. It turns out that
resonant terms cannot be systematically eliminated at fourth order, and as a consequence there is an undetermined constant
at third order. Therefore we look for a two-term generalised asymptotic solution of the form
\begin{equation}
	\label{eq:ansp2}
	\varphi(\tau)
	\sim
	\frac{\varphi_1(\sigma)}{\sigma}	 + \frac{\varphi_2(\sigma)}{\sigma^2}	 + o(1/\sigma^2),
\end{equation}
where
\begin{equation}
	\sigma = \tau^{1/2},
\end{equation}
$\varphi_1(\sigma)$ and $\varphi_2(\sigma)$ are periodic functions of $\sigma$.
By substituting the ansatz~(\ref{eq:ansp2}) into equation~(\ref{eq:mep2})
we find the following equations for $\varphi_1(\sigma)$ and $\varphi_2(\sigma)$, which are the analogues of
equations~(\ref{eq:p1f1}) and~(\ref{eq:p1f2}):
\begin{equation}
	\label{eq:p2f1}
	\varphi''_1 + 8 (\varphi_1)^3 = 0,
\end{equation}
\begin{equation}
	\label{eq:p2f2}
	\varphi''_2 + 24 (\varphi_1)^2 \varphi_2 = (3 - ( (\varphi'_1)^2 + 4 \varphi_1^4 )^{1/2})\varphi'_1.
\end{equation}
Equation~(\ref{eq:p2f1}) for $\varphi_1(\sigma)$ is a nonlinear differential equation whose solutions can be written
in terms of the Jacobi elliptic function $\mathrm{sn}$ of modulus $-1$~\cite{AS72}:
\begin{equation}
	\label{eq:sn}
	\varphi_1(\sigma) = b \sn( 2 b (\sigma-\sigma_0)|-1).
\end{equation}
By substituting this solution into equation~(\ref{eq:p2f2}) we find
\begin{equation}
	\label{eq:p2f2bis}
	\varphi''_2 + 24 (\varphi_1)^2 \varphi_2 = (3 - 2 b^2) \varphi'_1.
\end{equation}
Incidentally, that the square root $( (\varphi'_1)^2 + 4 \varphi_1^4 )^{1/2}$ in equation~(\ref{eq:p2f2})
reduces to the constant $2b^2$ is just the law of energy conservation for equation~(\ref{eq:p2f1}) considered as a
one-dimensional dynamical system. This fact, which also follows from the explicit form of the
solution~(\ref{eq:sn}), will be essential in our discussion of the general case in the next section.
Equation~(\ref{eq:p2f2bis}) is a nonhomogeneous linear differential equation with a non-constant coefficient.
However, it is immediate to find two linearly independent solutions of the homogenous equation, namely
$u_1(\sigma)=\varphi'_1(\sigma)$ and $u_2(\sigma)=\varphi_1(\sigma) + \sigma \varphi'_1(\sigma)$, and
using the method of variation of constants, to find the general solution:
\begin{equation}
\fl
	\varphi_2(\sigma)
	=
	c \varphi'_1(\sigma) + d (\varphi_1(\sigma) + \sigma \varphi'_1(\sigma))
	+
	\frac{3-2b^2}{3} \left( \sigma\varphi_1(\sigma)  + \frac{\sigma^2}{2} \varphi'_1(\sigma)\right).
\end{equation}
The condition that $\varphi_2(\sigma)$ be bounded requires that
\begin{equation}
	d = 0, \quad b = \sqrt{\frac{3}{2}},
\end{equation}
and we obtain explicit formulas for the coefficients $\varphi_1(\sigma)$ and $\varphi_2(\sigma)$ in the generalised
asymptotic solution~(\ref{eq:ansp2}), 
\begin{equation}
	\label{eq:32}
	\varphi_1(\sigma) = \sqrt{\frac{3}{2}} \sn(\sqrt{6}(\sigma-\sigma_0)|-1),
\end{equation}
\begin{eqnarray}
	\varphi_2(\sigma)  & = c \varphi'_1(\sigma) \\
	                              & = 3 c \cn(\sqrt{6}(\sigma-\sigma_0)|-1)\dn(\sqrt{6}(\sigma-\sigma_0)|-1),
\end{eqnarray}
where $\mathrm{cn}$ and $\mathrm{dn}$ denote Jacobi elliptic functions~\cite{AS72}. Note that again the
generalised asymptotic solution depends on two constants $\sigma_0$ and $c$, and that since
$\sigma=\tau^{1/2} = (mt)^{1/2}$, the frequency of the oscillations is not constant.

\begin{figure}
	 \begin{center}
		\includegraphics[width=10cm]{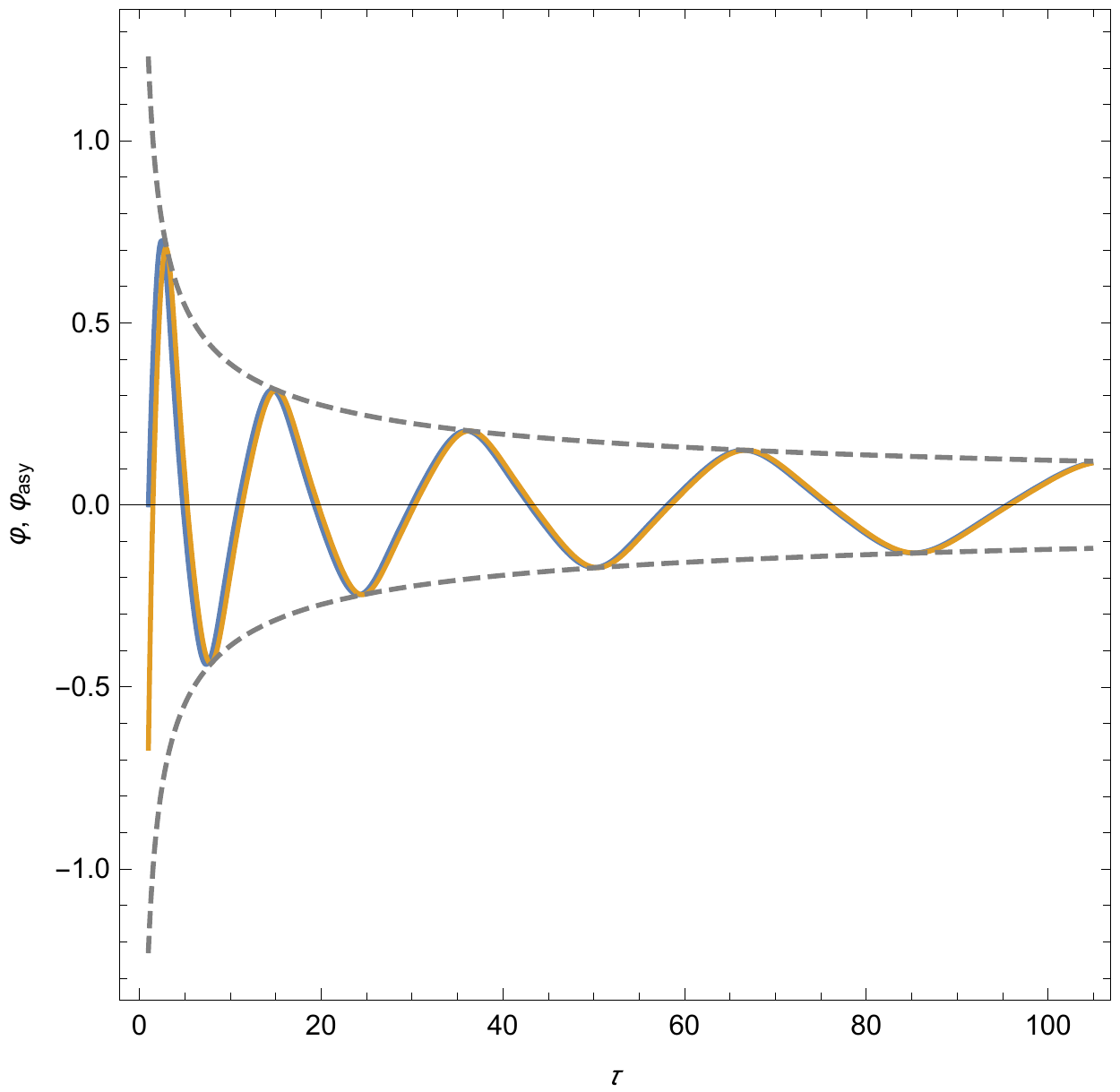}
	\end{center}
	\caption{Numerical integration (blue line) with initial conditions $\varphi(1)=0$, $\varphi'(1)=1$ and leading term
	              of the asymptotic expansion (brown line) for the differential equation~(\ref{eq:mep2}) corresponding
	              to the quartic potential. The dashed lines are the envelopes $\pm (3/(2\tau))^{1/2}$ that
	              set the scale for the generalised asymptotic solution~(\ref{eq:ansp2}).\label{fig:v4s}}
\end{figure}

In figure~\ref{fig:v4s} we plot the results of  a numerical integration of the the differential equation~(\ref{eq:mep2})
with initial conditions $\varphi(1)=0$, $\varphi'(1)=1$,
as well as the leading term of the asymptotic expansion~(\ref{eq:ansp2}). In this case,
the dashed lines are the envelopes $\pm (3/(2\tau))^{1/2}$ that
set the scale for the generalised asymptotic expansion~(\ref{eq:ansp2}), and
the factor $(3/2)^{1/2}$ is the prefactor in equation~(\ref{eq:32}) for $\varphi_1(\tau)$.

\begin{figure}
	 \begin{center}
		\includegraphics[width=10cm]{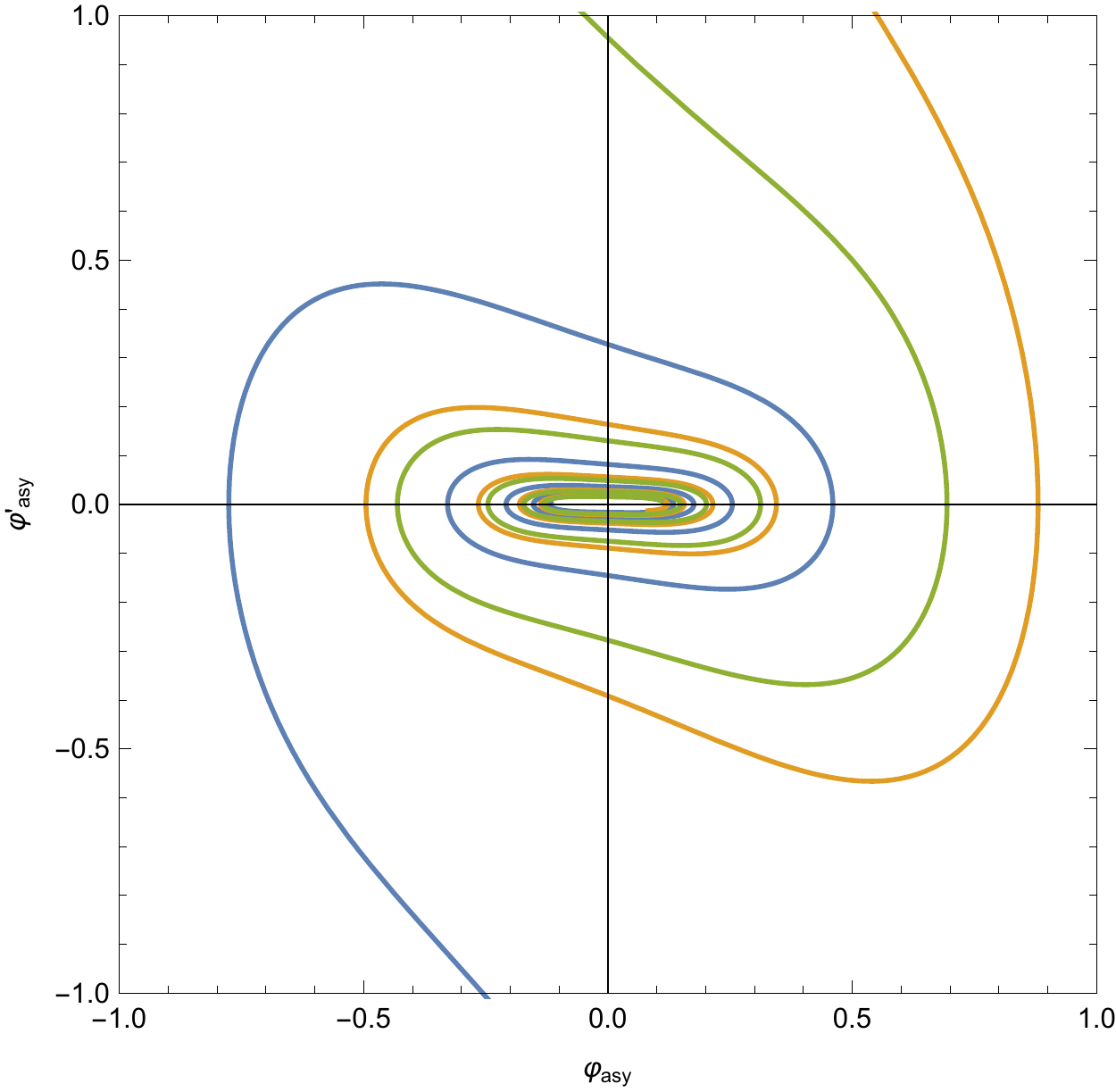}
	\end{center}
	\caption{Three typical asymptotic trajectories on the phase map of the differential equation~(\ref{eq:mep2})
	              corresponding to the quartic potential. Note that even for arbitrarily large times the approximation
	              of the representative point to the origin is not monotonic.\label{fig:v4p}}
\end{figure}

As in the quadratic case, in figure~\ref{fig:v4p} we plot three typical asymptotic trajectories on the phase
map of the differential equation~(\ref{eq:mep2}). Note that even for arbitrarily large times the approximation
of the representative point to the origin is not monotonic. (In this case $h(\tau)$ does not admit the same
geometric interpretation as in the quadratic case.) 

Again, using equations~(\ref{eq:ht}) and~(\ref{eq:htau}) we find the asymptotic formula for the Hubble parameter,
\begin{equation}
	H(t) \sim \frac{1}{2 t} - \frac{3m}{(6 m t)^{3/2}} \sn( s(t) |-1)
	                                                                         \cn( s(t) |-1)
	                                                                         \dn( s(t) |-1) + \cdots,
\end{equation}
where
\begin{equation}
	s(t) = \sqrt{6 m t} - \sqrt{6 m t_0},
\end{equation}
and using equation~(\ref{esp}) the corresponding formula for the eos parameter, 
\begin{equation}
	w_\Phi(t) \sim 1 - 2 \sn( s(t) |-1)^4+\cdots.
	\label{eq:wf4t}
\end{equation}
Note that at this order $w_\Phi(t)$ is independent of $c$. In the following section we show that the leading
term of the usual averages of this eos parameter is $1/3$.
\section{The general case\label{sec:vp}}
In this section we discuss the generalised asymptotic solution of equation~(\ref{eq:me}) for any value of $p$.
The results of this section include as particular cases the quadratic potential $(p=1)$ and the quartic potential $(p=2)$
studied in the previous sections, but the results obtained there were sharper than the results we will obtain in this section.
In the quadratic case we could find explicit formulas and a complete asymptotic expansion.
In the quartic case we could still find explicit formulas although we had to limit ourselves to a two-term asymptotic solution.
In this section $\varphi_1(\sigma)$ will be defined as the solution of an implicit equation.

The general equation, which we repeat here for convenience, is 
\begin{equation}
	\label{eq:mep}
	\varphi''(\tau) + ( \varphi'(\tau)^2 + \varphi(\tau)^{2p} )^{1/2} \varphi'(\tau) + p \varphi(\tau)^{2p-1} = 0.
\end{equation}
In analogy with the quartic case, we look for a two-term generalised asymptotic solution of the form
\begin{equation}
	\label{eq:ansp}
	\varphi(\tau)
	\sim
	\frac{\varphi_1(\sigma)}{\sigma}	 + \frac{\varphi_2(\sigma)}{\sigma^2}	 + o(1/\sigma^2),
\end{equation}
where
\begin{equation}
	\sigma = \tau^{1/p},
\end{equation}
and the conditions on the coefficients are the same as in the previous section. By substituting the ansatz~(\ref{eq:ansp})
into equation~(\ref{eq:mep}) we find the following equations for $\varphi_1$ and $\varphi_2$:
\begin{equation}
	\label{eq:pf1}
	\varphi''_1 + p^3 (\varphi_1)^{2p-1} = 0,
\end{equation}
\begin{equation}
	\label{eq:pf2}
	\varphi''_2 + p^3(2p-1) (\varphi_1)^{2p-2} \varphi_2 = (p+1 - ( (\varphi'_1)^2 + p^2 \varphi_1^{2p} )^{1/2})\varphi'_1.
\end{equation}
(Again, primes denote derivatives of the functions with respect to their arguments, in this case $\sigma$.)

In general, equation~(\ref{eq:pf1}) cannot be solved in closed form. However, it is readily
interpreted as the Newton's equation of motion with respect to the time $\sigma$ of a unit-mass
particle with position $\varphi_1$ under the confining potential $U(\varphi_1) = p^2 (\varphi_1)^{2p}/2$.
The corresponding conserved energy is
\begin{equation}
	\label{eq:e}
	E = \frac{1}{2} (\varphi'_1)^2 + \frac{p^2}{2} (\varphi_1)^{2p},
\end{equation}
and the solution $\varphi_1(\sigma)$ is implicitly defined by
\begin{equation}
	\label{eq:idef}
	\sigma - \sigma_0=\int_{\varphi_1^{(0)}}^{\varphi_1}\frac{\rmd x}{\sqrt{2E-p^2 x^{2p}}},
\end{equation}
or, equivalently, by
\begin{equation}
	\sigma - \sigma_0 = \frac{\varphi_1}{\sqrt{2E}}
	                                        \,
	                                       {}_2\mathrm{F}_1\!\left(
	                                       \frac{1}{2},
	                                       \frac{1}{2p},
	                                       1+\frac{1}{2p},
	                                       \frac{p^2}{2E} (\varphi_1)^{2p}
	                                       \right),
\end{equation}
where ${}_2\mathrm{F}_1$ denotes Gauss' hypergeometric function. Therefore $\varphi_1(\sigma)$
represents a periodic motion on the interval $-(2E/p^2)^{1/2p}\leq \varphi_1\leq (2E/p^2)^{1/2p}$.

Equation~(\ref{eq:pf2}) is again a nonhomogeneous linear differential equation with a non-constant coefficient.
Two linearly independent solutions of the homogenous equation are
$u_1(\sigma)=\varphi'_1(\sigma)$ and $u_2(\sigma)=\varphi_1(\sigma) + (p-1) \sigma \varphi'_1(\sigma)$, and
using the method of variation of constants we find the general solution for $\varphi_2(\sigma)$ in terms of
$\varphi_1(\sigma)$, namely
\begin{eqnarray}
	\varphi_2(\sigma) = & c \varphi'_1(\sigma) + d ( \varphi_1(\sigma) + (p-1) \sigma \varphi'_1 (\sigma))
	\nonumber\\
	   &
	      {}+ \left(1-\frac{\sqrt{2E}}{p+1}\right)
	                    \left(\sigma \varphi_1(\sigma) + \frac{p-1}{2} \sigma^2 \varphi'_1(\sigma)\right).
\end{eqnarray}
And the condition that $\varphi_2(\sigma)$ be bounded leads to
\begin{equation}
	\label{eq:ep}
	d = 0, \quad E = \frac{1}{2} (p+1)^2,
\end{equation}
and to
\begin{equation}
	\label{eq:phi2p}
	\varphi_2(\sigma) = c \varphi'_1(\sigma).
\end{equation}

We may also determine the first terms of the expansion of the reduced Hubble parameter~(\ref{eq:htau}),
\begin{equation}
	\label{nh1}
	h(\tau) \sim \frac{h_1(\sigma)}{\sigma^p}+\frac{h_2(\sigma)}{\sigma^{p+1}}+o(1/\sigma^{p+1}),
\end{equation}
where
\begin{equation}
	h_1(\sigma) = \frac{1}{p} (\varphi'_1(\sigma)^2+p^2\varphi(\sigma)^{2p})^{1/2},
\end{equation}
\begin{equation}
	h_2(\sigma) = \frac{1}{p^2 h_1(\sigma)}
	( \varphi'_1(\sigma)\varphi'_2(\sigma) + p^3 \varphi_1(\sigma)^{2p-1} \varphi_2(\sigma) - \varphi_1(\sigma)\varphi'_1(\sigma)).
\end{equation}
As a consequence of (\ref{eq:pf1}), (\ref{eq:e}), (\ref{eq:ep}) and (\ref{eq:phi2p}) we have
\begin{eqnarray}
	\label{eq:hp1}
	h_1(\sigma)=\frac{p+1}{p},\\
	\label{eq:hp2}
	h_2(\sigma)=-\frac{1}{p(p+1)} \varphi_1(\sigma) \varphi'_1(\sigma).
\end{eqnarray}
Then the leading term of the eos parameter is
\begin{eqnarray}
	\label{eosp}
	w_{\Phi}(t) & \sim \frac{p-1}{p+1}-\frac{2p}{(p+1)^2}h_2'(\sigma)
	\\
	 \label{eosp2}
	                  & =1-\frac{2p^2}{(p+1)^2} \varphi_1(\sigma)^{2p}.
\end{eqnarray}
In particular, the first terms of equation~(\ref{eq:wf2t}) and of equation~(\ref{eq:wf4t}) are respectively the cases $p=1$ and $p=2$ of equation~(\ref{eosp2}).
That at this order $w_\Phi(t)$ is independent of $c$ is thus a general result.

As we mentioned in the introduction, different averaging procedures lead to the same leading contribution of the eos parameter~(\ref{eq:winf}).
Equation~(\ref{eosp}) leads to the same result if we average over a period of the variable $\sigma$ as an immediate consequence
of this equation and of the periodicity of $h_2(\sigma)$ in $\sigma$ (see equation~(\ref{eq:hp2})):
\begin{equation}
	\label{aver}
	\langle w_{\Phi} \rangle_1=\frac{1}{T_\sigma}\int_0^{T_\sigma} w_{\Phi}(\sigma) d \sigma\sim \frac{p-1}{p+1}.
\end{equation}
It is also easy to derive the same result for other averages used in the literature. For example, Rendall~\cite{RE07}
uses the averaging procedure defined by
\begin{equation}
	\label{aver1}
	\langle w_{\Phi} \rangle_2
	=
	\frac{\displaystyle\int_t^{\infty}p_{\Phi}(t)\rmd t}{\displaystyle\int_t^{\infty}\rho_{\Phi}(t)\rmd t}.
\end{equation}
Note that in the variable $\tau$,
\begin{eqnarray}
	\label{pro1}
	p_{\Phi}(\tau)     & = \frac{M^4}{p}\left( \frac{2}{3} \right)^p(\varphi'(\tau)^2-\varphi(\tau)^{2p}),\\
	\rho_{\Phi}(\tau) & = \frac{M^4}{p}\left( \frac{2}{3} \right)^p(\varphi'(\tau)^2+\varphi(\tau)^{2p}).
\end{eqnarray}
Using equation~(\ref{eq:hprimetau}) we find that,
\begin{equation}
	\label{c2}
	\int_{\tau}^{\infty} \varphi'(\tau)^2 d \tau=h(\tau)\sim \frac{p+1}{p}\frac{1}{\tau},
\end{equation}
where the asymptotic estimate follows from equations~(\ref{nh1}) and~(\ref{eq:hp1}).
Furthermore, using the following consequence of equations~(\ref{eq:me}) and~(\ref{eq:htau}),
\begin{equation}
	\label{con}
	p\varphi^{2p}(\tau)
	=
	-h'(\tau)
	-\frac{1}{2}\varphi(\tau)^2 \varphi'(\tau)^2
	-\frac{\rmd}{\rmd\tau} \left(\varphi (\tau) \varphi'(\tau) +\frac{1}{2}h(\tau) \varphi(\tau)^2\right),
\end{equation}
we note that as $\tau \rightarrow \infty$
\begin{equation}
	h(\tau) \left( 1+\frac{1}{2}\varphi^{2}(\tau) \right) =O(1/\tau),
\end{equation}
\begin{equation}
	\varphi(\tau) \varphi'(\tau) = O(1/\tau^{1+1/p}),
\end{equation}
and
\begin{equation}
	\int_{\tau}^{\infty}\varphi(\tau)^2\varphi'(\tau)^2 d \tau = O(1/\tau^{1+2/p}).
\end{equation}
Therefore
\begin{equation}
	\label{c3}
	\int_{\tau}^{\infty}\varphi^{2p(\tau)} d \tau \sim \frac{p+1}{p^2} \frac{1}{\tau},
\end{equation}
and from~(\ref{c2}) and~(\ref{c3}) we obtain that indeed
\begin{equation}
	\langle w_{\Phi} \rangle_2 \sim \frac{p-1}{p+1}.
\end{equation}
\section{Concluding remarks \label{sec:con }}
We have analysed the asymptotic properties of the dynamical equation of the inflaton during the oscillatory phase of
reheating to extend the existing results for the quadratic case,  to provide new explicit results for the quartic case,
and implicit results (that include the former as particular cases) for potentials that close
to their global minima behave as even monomial potentials.

We have paid special attention to the derivation of explicit expressions for the instantaneous eos parameter $w_{\Phi}$,
whose averages turn out to be in agreement with the averaged values found in the literature.
The effect of the difference between the instantaneous
value $w_\Phi(t)$ and the averaged value $\langle w_\Phi \rangle$ in equation~(\ref{eq:winf})
may be of interest in realistic models due to the finite duration of reheating~\cite{MA10,CO15,EL15,UE16,DR17}.
Furthermore, as pointed out by Mukhanov~\cite{MU05}, although observationally the corrections to the instantaneous
Hubble parameter $H(t)$ may be negligible, they must be taken into account when calculating curvature invariants.

Finally, as a line of future development, we mention the possible existence of a finer scale than $\tau^{1/p}$ that might permit
the calculation of full asymptotic expansions for the inflaton, instead of the truncated asymptotic formulas derived in
the present paper for the degenerate potentials.
 \section*{Appendix A\label{sec:appa}}
In this appendix we prove that the ansatz~(\ref{eq:fn}) indeed leads to a recurrence procedure that can
be carried out to arbitrary order. To this end, first note that
(for any $p$) the main equation~(\ref{eq:me}) and the definition~(\ref{eq:htau}) for $h(\tau)$ imply that
\begin{eqnarray}
	\label{eq:sys1}
	\varphi''(\tau) + h(\tau) \varphi'(\tau) + p \varphi(\tau)^{2p-1} = 0,\\
	\label{eq:sys2}
	h'(\tau) = -\varphi'(\tau)^2.
\end{eqnarray}
Conversely, since the system~(\ref{eq:sys1})--(\ref{eq:sys2}) implies that
\begin{equation}
	\frac{\rmd}{\rmd\tau}\left( h^2 - ((\varphi')^2 + \varphi^{2p}) \right) = 0,
\end{equation}
and both $\varphi$ and $\varphi'$ tend to zero, it follows that the system (\ref{eq:sys1})--(\ref{eq:sys2}) and the
main equation~(\ref{eq:me}) are equivalent~\cite{RE07}.

Hereafter we set $p=1$. In addition to the expansions~(\ref{eq:ansp1}) and (\ref{eq:fn}) for $\varphi(\tau)$,
and~(\ref{eq:htaup1}) for $h(\tau)$, it is convenient to introduce an expansion for $\varphi'(\tau)$,
\begin{equation}
	\label{phip}
	\varphi'(\tau) = \sum_{n=1}^\infty \frac{\psi_n(\tau)}{\tau^{n}},
\end{equation}
where
\begin{equation}
	\psi_n(\tau) = \varphi'_n(\tau) - (n-1)\varphi_{n-1}(\tau),
\end{equation}
and we set $\varphi_0(\tau)\equiv 0$. Then, the system~(\ref{eq:sys1})--(\ref{eq:sys2}) leads to
the following system of differential equations:
\begin{eqnarray}
	\label{sys2p}
	\varphi''_n + \varphi_n=(n-1) ( 2\varphi'_{n-1} - (n-2) \varphi_{n-2}) - \sum_{l=1}^{n-1}\psi_l h_{n-l},\\
	\label{sys2h}
	h'_n = (n-1) h_{n-1} - \sum_{l=1}^{n-1}\psi_l \psi_{n-l}.
\end{eqnarray}

For $n=1$ these equations are
\begin{equation}
	\label{hom} 
	\varphi''_1 + \varphi_1 = 0,
\end{equation}
\begin{equation}
	h'_1 = 0,
\end{equation}
and their general solutions are
\begin{equation}
	\label{hom1} 
	\varphi_1(\tau) = c_{1,1}\cos(\tau-\tau_0),
\end{equation}
\begin{equation}
	h_1(\tau) = f_{1,0},
\end{equation}
respectively, where $c_{1,1}$ and $f_{1,0}$ are arbitrary real constants.  

For $n\geq 2$ the  solution of~(\ref{sys2p})--(\ref{sys2h}) can be written as
\begin{equation}
	\label{sollin}
	\varphi_n(\tau) = c_{n,1}\cos(\tau-\tau_0) + d_{n,1}\sin(\tau-\tau_0) + \varphi_n^{(p)}(\tau),
\end{equation}
\begin{equation}
	h_n(t) = f_{n,0} + h_{n}^{(p)}(\tau),
\end{equation}
where $\varphi_n^{(p)}(\tau)$ and $h_{n}^{(p)}(\tau)$ are particular solutions of~(\ref{sys2p}) and~(\ref{sys2h}) respectively.
To avoid secular terms in the functions  $\varphi_n^{(p)}(\tau)$ and $h_{n}^{(p)}(\tau)$ we have to ensure that the inhomogeneous
part of~(\ref{sys2p}) is free of terms proportional to $\cos(\tau-\tau_0)$ or $\sin(\tau-\tau_0)$,
and that the inhomogeneous part of~(\ref{sys2h}) does not contain a constant term.

We now prove that this process of eliminating secular terms yields an expansion~(\ref{eq:ansp1}) for $\varphi(\tau)$
with coefficients of the form~(\ref{eq:fn}), and that the corresponding expansion~(\ref{eq:htaup1}) is of the form
\begin{equation}
	\label{eq:hn}
 	h_n(\tau)=\sum_{k=0,k\,\mathrm{even}}^{n} f_{n,k}\cos (k(\tau-\tau_0))+g_{n,k}\sin(k(\tau-\tau_0)),
\end{equation}
with $f_{n,n}=0$ for even $n$.

Thus, for $n=2$ the equations~(\ref{sys2p}) and~(\ref{sys2h}) are
\begin{equation}
	\label{hom2} 
	\varphi''_2+\varphi_2=(2-h_1) \varphi'_1
\end{equation}
\begin{equation}
	\label{hom2h}
	h'_2=h_1-(\varphi'_1)^2.
\end{equation}
Then, taking into account~(\ref{hom1}), the inhomogeneous parts of~(\ref{hom2}) and~(\ref{hom2h}) become
\begin{equation}
	\label{hom21}
	(2-h_1(\tau)) \varphi'_1(\tau) = (f_{1,0}-2) c_{1,1} \sin(\tau-\tau_0),
\end{equation}
\begin{equation}
	h_1(\tau) - \varphi'_1(\tau)^2 = f_{1,0}-\frac{c_{1,1}^2}{2}+\frac{c_{1,1}^2}{2}\cos(2(\tau-\tau_0)).
 \end{equation}
Therefore, the particular solutions $\varphi_2^{(p)}(\tau)$ and $h_2^{(p)}(\tau)$ of~(\ref{hom2}) are free of resonant terms provided
\begin{equation}
	\label{res1}
	f_{1,0}=2,\quad c_{1,1}=2.
\end{equation}
Then we obtain
\begin{equation}
	\label{hom1o} 
	\varphi_1(\tau)=2\cos(\tau-\tau_0),
\end{equation}
\begin{equation}
	h_1(\tau)=2,
\end{equation}
and
\begin{equation}
	\label{hom22} 
	\varphi_2(\tau) = c_{2,1}\cos(\tau-\tau_0) + d_{2,1}\sin(\tau-\tau_0),
\end{equation}
\begin{equation}
	h_2(t) = f_{2,0} + \sin(2(\tau-\tau_0)).
\end{equation}
The same procedure for $n=3$ leads to
\begin{equation}
	c_{2,1} = f_{2,0} = c,
	\quad
	d_{2,1} = - \frac{1}{2},
\end{equation}
where $c$ is a free parameter.

Now we proceed by induction and assume that for any given integer $n\geq 3$ we have already obtained
the coefficients  $\varphi_k$ and $h_k$ with $k=1,\ldots,n-1,$ of the form~(\ref{eq:fn}) and~(\ref{eq:hn}).

The secular terms in equation~(\ref{sys2h}) for $h_n$ come from
\begin{equation}
	(n-1)h_{n-1}-2 \psi_1\psi_{n-1}-\sum_{l=2}^{n-2} \psi_l \psi_{n-l},
\end{equation}
and vanish provided that
\begin{equation}
	\label{sec1}
	-2c_{n-1,1} + (n-1) f_{n-1,0}=R_{n-2},
\end{equation}
where $R_{n-2}$ denotes a sum of terms depending on the coefficients $c_{k,j}$ $d_{k,j}$
with $k\leq n-2$.

Because of our induction hypothesis, from the structure of the right-hand side of the equation~(\ref{sys2h})
it is straightforward to deduce  that $h_{n}$ is  also  of  the form~(\ref{eq:hn}). For example, if $n$ is odd ($n=2r+1$),
then  $h_{n-1}$ is a linear combination of $\cos(2k(\tau-\tau_0))$ and $\sin(2k'(\tau-\tau_0))$
with $k=0,1,\ldots, r-1$ and $k'=1,\ldots,r$.  Similarly, the  products $\psi_l \psi_{n-l}$ with $l=1\ldots 2r$  in
equation~(\ref{sys2h}) are linear combinations of $\cos(2k(\tau-\tau_0))$ and $\sin(2k(\tau-\tau_0))$ with $k=0,1,\ldots, r$.
Therefore $h_{n}$ is of the form~(\ref{eq:hn}).

Next we note that the secular terms in  equation~(\ref{sys2p}) for $\varphi_n$ come from the terms
\begin{equation}
	(n-1)\left(2 \varphi'_{n-1}-(n-2) \varphi_{n-2}\right) -\sum_{l=1}^{n-1} \psi_l h_{n-l},
\end{equation}
and are given by
\begin{eqnarray}
 	\label{sec11}
	-2(n-2)\left(c_{n-1,1}\sin (\tau-\tau_0)-d_{n-1,1}\cos(\tau-\tau_0)\right) \nonumber \\
	\quad{}+ 2 f_{n-1,0}\,\sin(\tau-\tau_0) -S_{n-2}\sin (\tau-\tau_0)-T_{n-2}\cos(\tau-\tau_0),
\end{eqnarray}
 where $S_{n-2}$ and $T_{n-2}$ denote sums of terms depending on the coefficients $c_{k,j}$, $d_{k,j}$,
$f_{k,j}$, $g_{k,j}$, with $k\leq n-2$ and on $f_{n-1,2}$, $g_{n-1,2}$.
Therefore, to eliminate the secular terms in $\cos (\tau-\tau_0)$ and $\sin(\tau-\tau_0)$ we must set
\begin{equation}
	\label{sec13}
 	2(n-2)\, d_{n-1,1}=T_{n-2},
\end{equation}
and
\begin{equation}
	\label{sec12}
	-2(n-2)c_{n-1,1}+2 f_{n-1,0}=S_{n-2},
\end{equation}
respectively.
Equation~(\ref{sec13}) determines $d_{n-1,1}$ in terms of the coefficients $c_{k,j}$ $d_{k,j}$, $f_{k,j}$, $g_{k,j}$with $k\leq n-2$
and on $g_{n-1,2}$.
Furthermore,  equations~(\ref{sec1}) and~(\ref{sec12}) are a linear system  for $c_{n-1,1}$ and $f_{n-1,0}$ whose
coefficient matrix has determinant $2n(n-3)$. Thus, this linear system is undetermined for $n=3$ (with a free unknown
$c=c_{2,1}=f_{2,0}$) and it has a unique solution  for $n>3$.
In this way, it is clear that all the coefficients $c_{n,1}$, $d_{n,1}$ and $f_{n,0}$, with the exception of
$c_{2,1}=f_{2,0}$, are determined  in terms of $c$. 

A similar argument proves the ansatz~(\ref{eq:fn}) for $\varphi_n$.
\section*{Acknowledgments}
The financial support of the Spanish Ministerio de Econom\'{\i}a y
Competitividad under Project No. FIS2015-63966-P is gratefully acknowledged.
\section*{References}
\bibliographystyle{iopart-num}
\bibliography{inflation-g}
\end{document}